\documentclass[twocolumn,pra,amsmath,amssymb,showpacs,nofootinbib]{revtex4}
\usepackage{graphicx}
\usepackage{dcolumn}
\usepackage{bm}
\usepackage{latexsym}
\usepackage{amstext}
\usepackage{amsxtra}

\newcommand{\kb}[0]{k_{\mathrm{B}}}
\newcommand{\eps}[0]{\varepsilon}
\newcommand{\moy}[1]{\left \langle #1 \right \rangle}

\begin{document}
\title{Kinetics of the evaporative cooling of an atomic beam}
\author{Thierry Lahaye}
\author{David Gu\'ery-Odelin}
\affiliation{Laboratoire Kastler Brossel$^{*}$, 24 rue Lhomond,
F-75231 Paris Cedex 05, France}
 \date{\today}

\begin{abstract}
We compare two distincts models of evaporative cooling of a
magnetically guided atomic beam: a continuous one, consisting in
approximating the atomic distribution function by a truncated
equilibrium distribution, and a discrete-step one, in which the
evaporation process is described in terms of successive steps
consisting in a truncation of the distribution followed by
rethermalization. Calculations are performed for the semi-linear
potential relevant for experiments. We show that it is possible to
map one model onto the other, allowing us to infer, for the
discrete-step model, the rethermalization kinetics, which turns
out to be strongly dependent upon the shape of the confining
potential.
\end{abstract}


\maketitle

\section{Introduction}

Evaporative cooling~\cite{KetterleVanDruten} is a very powerful
technique that allowed the achievement of quantum degeneracy in
dilute atomic vapors~\cite{bec}. On the theoretical side, apart
from direct numerical simulations~\cite{wu96}, several models of
evaporative cooling have been studied, which can be classified in
two categories: \emph{continuous} and \emph{discrete} ones.

In discrete models, the process of evaporative cooling is
approximated as a series of truncations of the atomic distribution
function, followed by rethermalization towards
equilibrium~\cite{ketterleEvap}. Through many steps, the
phase-space density of the atomic sample increases up to order
unity. The advantage of such models lies in their simplicity.
However, in the case of trapped clouds of atoms, those models are
not realistic, since, experimentally, the evaporation is done by
ramping down \emph{continuously} a radio-frequency knife.
Moreover, such models give no indication on the kinetics of the
evaporation. Therefore, one needs to resort to more elaborated,
continuous models, using an appropriate Ansatz for the
distribution function, namely a truncated equilibrium function. It
is then possible, starting from Boltzmann equation, to obtain the
time evolution of the temperature and of the number of
atoms~\cite{walraven}. Quantitative comparisons between the
predictions of those two types of models have not been made to
date.

Evaporative cooling has been revisited in the context of the
cooling of a guided atomic beam, in view of achieving a
continuous-wave atom laser. The proposal~\cite{Mandonnet00} used a
continuous evaporation model to predict the possibility of
achieving quantum degeneracy by using transverse evaporation on an
atomic beam, confined transversally by an harmonic potential. This
proposal triggered experimental work in several groups, and a
discrete model of the evaporative cooling of a beam was
developed~\cite{epjd} in close connection with recent
experiments~\cite{prlrb2,lahaye05}.

In this paper, we address the problem of the transverse
evaporative cooling of a magnetically guided atomic beam with an
energy-selective ``knife''. We first develop a continuous model
using hydrodynamic equations, adapted from~\cite{Mandonnet00}, for
an experimentally realistic transverse potential. Then we describe
the same process with a discrete-step evaporation model, analog to
the one used in~\cite{epjd}. Finally, we compare the results given
by those two distincts models. In particular, this comparison
allows us to study the influence of the shape of the confining
potential on the kinetics of rethermalization and on the
evaporation `ramp'.

We consider an atomic beam with a flux $\Phi$, a mean longitudinal
velocity $v$, and a temperature $T$, propagating in a quadrupole
magnetic guide of axis~$z$, providing the following semi-linear
transverse potential~:
\begin{equation}
U(r)=\mu\sqrt{B_0^2+b^2 r^2}-\mu B_0\,. \label{Eq:1}
\end{equation}
Here, $\mu$ is the magnetic moment of the atoms, $b$ the
transverse gradient of the two-dimensional quadrupole magnetic
field created by the guide, $r=\sqrt{x^2+y^2}$ the distance from
the guide axis, and $B_0$ a longitudinal bias field used to avoid
Majorana spin-flips~\cite{lahaye05}. The on-axis potential is
taken as the origin of energies. One defines the dimensionless
parameter $\alpha\equiv \mu B_0/\kb T$. For typical experimental
parameters, the evaporation starts with $\alpha\ll1$ where the
potential experienced by the atoms is essentially linear, and
degeneracy is reached in the regime $\alpha\gg1$, where the
potential is essentially harmonic. Therefore it is crucial to take
into account the real shape of the potential in order to describe
the whole evaporation process.

An important quantity characterizing the beam is the $s$-wave
elastic collision rate~$\gamma$, given by:
\begin{equation}
\gamma=\frac{\sigma}{2\pi^{3/2}}\,\frac{1+2\alpha}{(1+\alpha)^2}\,
\frac{\Phi}{v}\,\left(\frac{\mu
b}{\kb T}\right)^2\sqrt{\frac{\kb T}{m}}\,, \label{Eq:guide:cr}
\end{equation}
where $\sigma$ is the $s$-wave scattering cross-section.

Two-dimensional transverse evaporation is applied in order to
selectively remove atoms from the beam. In practice, this is
achieved by driving transitions to an untrapped state with
radio-frequency or microwave fields~\cite{lahaye05}. The
evaporation criterion then relates to the transverse energy $\eps$
\emph{and} to the angular momentum of the atom along $z$. For the
sake of simplicity, as in Ref.~\cite{walraven}, we assume in the
following that any atom having a transverse energy
$\eps\geqslant\eta\kb T$, where $\eta$ is the \emph{evaporation
parameter}, is evaporated. This criterion implicitly assumes
sufficient ergodicity of the atomic motion. Experimentally, this
simple energy criterion can be well approximated by the
multi-radii evaporation scheme described in~\cite{epjd}.

\section{Continuous model}

In this section, we assume that the evaporation takes place over
the whole guide length. The height of the energy knife $\eps_{\rm
ev}(z)=\eta\kb T$ depends on $z$ in order to perform \emph{forced}
evaporation. We therefore assume that the beam's distribution
function is a local equilibrium one~\cite{walraven}, truncated at
the energy $\eps_{\rm ev}$.

By using such an Ansatz in the Boltzmann equation, one gets a set
of coupled hydrodynamic equations relating the following local
quantities characterizing the beam: the linear density $n(z)$, the
longitudinal velocity $v(z)$, and the temperature $T(z)$. The
details of this somehow lengthy calculation can be found in
ref.~\cite{MandonnetThese,Mandonnet00} for the case of a harmonic
transverse potential. The validity of such an Ansatz was checked
with a molecular dynamics simulation of the process. In the
following, we present the generalization of this analytical
approach to the case of a semi-linear confinement.

In the stationary regime, the hydrodynamic equations read:
\begin{subequations}
\label{eq:hd}
\begin{eqnarray}
&&\!\!\!\!\!\!\!\!\!\!\!\!\!\!{\partial_z}(n v)=-\Gamma_1 n\,, \label{Eq:hd:1} \\
&&\!\!\!\!\!\!\!\!\!\!\!\!\!\!{\partial_z} (n v^2+n{v_{\rm
th}^2})=-\Gamma_1 n v\,, \label{Eq:hd:2}
\\
&&\!\!\!\!\!\!\!\!\!\!\!\!\!\!{\partial_z}\!\left[nv\!\!\left(\frac{5}{2}{v_{\rm
th}^2}+\frac{v^2}{2}+\frac{\moy{U}}{m}\right)\!\right]=-n\!\left(\Gamma_1\frac{v^2}{2}+\Gamma_2{v_{\rm
th}^2}\right)\!\!, \label{Eq:hd:3}
\end{eqnarray}
\end{subequations}
where $v_{\rm th}=\sqrt{\kb T/m}$. They correspond, respectively,
to the evolution of the flux, of the momentum, and of the enthalpy
of the beam. The notation $\moy{U}$ stands for the thermal
average, at the temperature $T$, of the potential energy $U$.
These equations are well suited to describe a supersonic beam with
a high enough Mach number (typically $v\gtrsim3v_{\rm th}$).

In the semi-linear potential (\ref{Eq:1}), the mean value of the
potential energy reads $\moy{U}=\kb T(2+\alpha)/(1+\alpha)$.
$\Gamma_1$ and $\Gamma_2$ correspond to the evaporation-induced
particle and energy loss rates, respectively. They are
proportional to the elastic collision rate $\gamma$ and obviously
depend on the evaporation parameter $\eta$:
\begin{equation}
\Gamma_{i}=\gamma
\sqrt{\frac{2}{\pi}}\frac{8}{15}K_{i}(\eta,\alpha)\qquad(i=1,2)\,.
\end{equation}
The functions $K_i$ are given by the following integral:
\begin{equation}
K_i(\eta,\alpha)=\int_{0}^{\eta+1/2}f(x,\alpha)g_{i}(x,\eta){\rm
d}x,\,
\end{equation}
with $f(x,\alpha)=x^{3/2}(5\alpha+2x){\rm
e}^{-\eta-1/2}/(1+2\alpha)$ being a contribution from the density
of states per unit length in the semi-linear potential,
$g_1(x,\eta)={\rm e}^{-x}(\eta-x-1/2)+{\rm e}^{-\eta-1/2}$, and
$g_2(x,\eta)={\rm e}^{-\eta-1/2}(3+2\eta-x)+{\rm
e}^{-x}[(\eta+1/2)^2-2-3x/2-\eta x]$.

\begin{figure}[t]
\includegraphics[width=8.5cm]{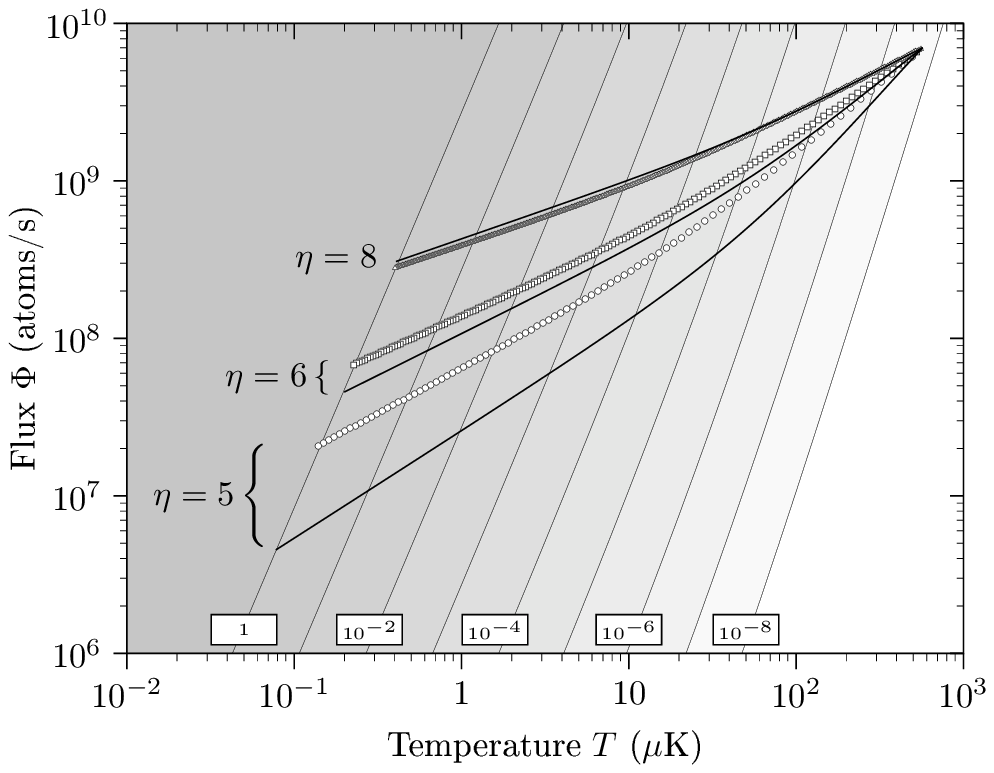}
\caption{Evaporation ``trajectories'' in the temperature/flux
plane, for evaporation with a constant $\eta$ for the continuous
model (solid lines) and for the discrete one (dots, each symbol
representing the effect of one evaporation step). The background
grey scale (with the white labels) shows the on-axis phase-space
density. The qualitative shape of those trajectories are similar
for both models. For~$\eta$ high enough (e.g., $\eta=8$), the
trajectories are very close in both models. The number of
evaporation zones required for reaching degeneracy in the
discrete-step model, for $\eta=5$ (resp. 6, 8), is 88 (resp. 152,
526). } \label{Fig:traj}
\end{figure}

The solid lines in Fig.~\ref{Fig:traj} depict the evolution of the
beam's flux and temperature obtained by solving the hydrodynamic
equations (\ref{eq:hd}), assuming that the evaporation parameter
$\eta$ remains constant throughout the evaporation. The initial
conditions are the experimental ones of Ref.~\cite{lahaye05}, in
which $^{87}{\rm Rb}$ ($\mu=\mu_{\rm B}/2$) atoms are used: at
$z=0$, one has $\Phi=7\times10^9$~at/s, $v=60$~cm/s, and
$T=570$~$\mu$K. The gradient is 800~G/cm and a $B_0=1$~G bias
field is applied. When one increases the value of $\eta$,
degeneracy (phase-space density $\rho\sim 1$) is achieved for
higher fluxes (and therefore higher temperatures) since the
evaporated particles are very energetic and consequently the
evaporative cooling is more efficient. The change in the slope of
the ``evaporation trajectories'' for $T\sim\mu
B_0/\kb\sim30$~$\mu$K is due to the fact that the confinement
experienced by the atoms changes from essentially linear
($\alpha=0.06\ll1$) to essentially harmonic ($\alpha\gg1$) as the
temperature is reduced. Indeed, the gain in phase-space density
scales differently with the shape of the
potential~\cite{ketterleEvap,walraven,epjd}.

\section{Discrete-step model}

We now turn to the description of the evaporative cooling process
with a discrete-step model. One evaporation step reduces the
atomic flux from $\Phi$ to $\Phi'$. After rethermalization, the
beam acquires a new temperature $T'$. In order to calculate the
relative variations of flux $\varphi=\Phi'/\Phi$ and of
temperature $\tau=T'/T$, we adapt the approach of Ref.~\cite{epjd}
to the semi-linear potential and to the energy-dependant
evaporation criterion. We therefore introduce the two-dimensional
density of states $\varrho(\eps)$ in the semi-linear
potential~(\ref{Eq:1}): $\varrho(\eps)\propto\eps[2+\eps/(\mu
B_0)]$. The fraction $\varphi$ of remaining atoms after one
evaporation step is:
\begin{eqnarray}
\varphi(\eta,\alpha)&=&\frac{\int_0^{\eta\kb
T}\varrho(\varepsilon)\,{\rm e}^{-\varepsilon/\kb T}\,{\rm
d}\varepsilon}{ \int_0^{\infty}\varrho(\varepsilon)\,{\rm
e}^{-\varepsilon/\kb T}\,{\rm d}\varepsilon}\\
&=&1-\frac{2+2\alpha(1+\eta)+\eta(2+\eta)}{2(1+\alpha)}\,{\rm
e}^{-\eta}\,.
\end{eqnarray}
In order to derive $\tau$, we first calculate the transverse
energy $\bar{\eps}$ of the remaining atoms:
\begin{equation}
\bar{\varepsilon}=\frac{ \int_0^{\eta\kb
T}\varepsilon\,\varrho(\varepsilon)\,{\rm e}^{-\varepsilon/\kb
T}\,{\rm d}\varepsilon}{ \int_0^{\infty}\varrho(\varepsilon)\,{\rm
e}^{-\varepsilon/\kb T}\,{\rm d}\varepsilon}\,.
\end{equation}
We define the dimensionless parameter $\theta(\eta,\alpha)\equiv
\bar{\eps}/(\kb T)$~\cite{theta}. The conservation of the total
energy during rethermalization gives:
\begin{equation}
\kb T\left(\theta\Phi+\frac{\Phi'}{2}\right) =%
\Phi'\moy{U}_{T',\alpha'}+\Phi'\frac{3\kb T'}{2}\,,
\end{equation}
where the average $\moy{U}_{T',\alpha'}$ of the potential is taken
at thermal equilibrium with a temperature $T'$, i.e. with
$\alpha'=\alpha/\tau$. This yields a quadratic equation in~$\tau$,
with the solution:
$$
\tau=\frac{2\theta\!+\varphi-5\alpha\varphi +\!{\sqrt{28 \varphi
\alpha(2   \theta + \varphi
)+{(2\theta+\varphi-5\alpha\varphi)^2}}}}{14 \varphi }.
$$
One then readily obtains the relative variations of the collision
rate $\gamma$ and of the phase-space density~$\rho$ after an
evaporation step.

The corresponding ``evaporation trajectories'' (for $\eta$
constant) are depicted with circles on Fig.~\ref{Fig:traj}. Each
symbol represents the flux and temperatures $(\Phi_n,T_n)$ of the
beam after the $n^{\rm th}$ evaporation zone.

\section{Discussion}

\begin{figure}[t!]
\includegraphics[width=7cm]{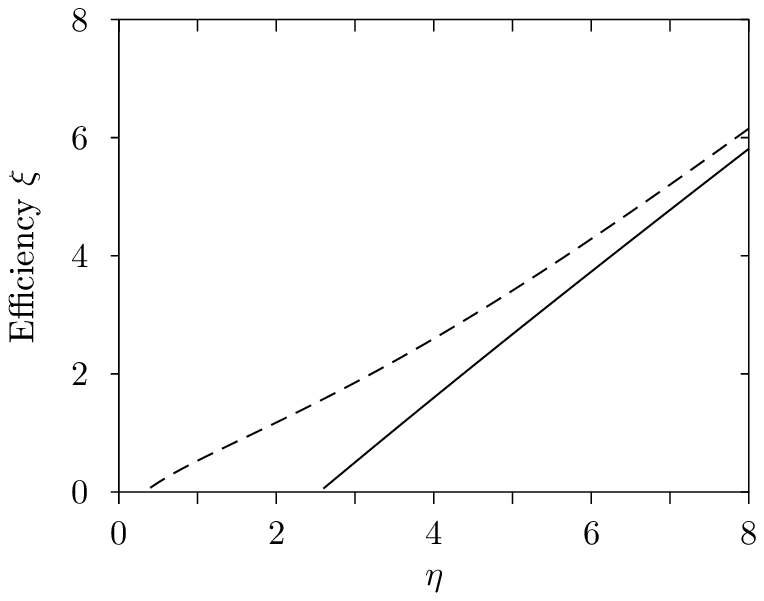}
\caption{Evaporation efficiency $\xi$ as a function of the
evaporation parameter $\eta$, for a harmonic transverse
confinement. The solid (resp. dashed) line corresponds to the
continuous (resp. discrete-step) evaporation model.}
\label{Fig:eff}
\end{figure}

We compare the results given by the two models, in terms of
evaporation trajectories and of the efficiency of evaporation. As
depicted on Fig.~\ref{Fig:traj}, the evaporation trajectories have
the same qualitative behavior in both models. However, for a given
$\eta$, the discrete-step evaporation leads to higher final fluxes
and temperatures. For a high evaporation parameter ($\eta=8$), the
trajectories given by both models almost coincide.

To make these statements more quantitative, we introduce the
figure of merit for an evaporative cooling ramp, i.e. the relative
variation of the beam's phase-space density $\rho$ for a given
loss of particles. We therefore define the evaporation
\emph{efficiency} $\xi$ as:
\begin{equation}
\xi\equiv-\frac{{\rm d}\ln \rho}{{\rm d}\ln\Phi}\,.
\end{equation}

This quantity is straightforward to calculate for a given
evaporation model. Fig.~\ref{Fig:eff} represents $\xi(\eta)$ for
the case of a harmonic transverse confinement. As expected, $\xi$
increases with $\eta$ in both models. It appears that
discrete-step evaporation is more efficient than the continuous
one, which can be understood qualitatively by the fact that in the
latter scheme, some atoms are evaporated without giving rise to a
temperature reduction, a process commonly called
``spilling''~\cite{KetterleVanDruten,walraven}. However, for
$\eta$ high enough, the efficiencies of both models almost
coincide.

\begin{figure}
\includegraphics[width=7cm]{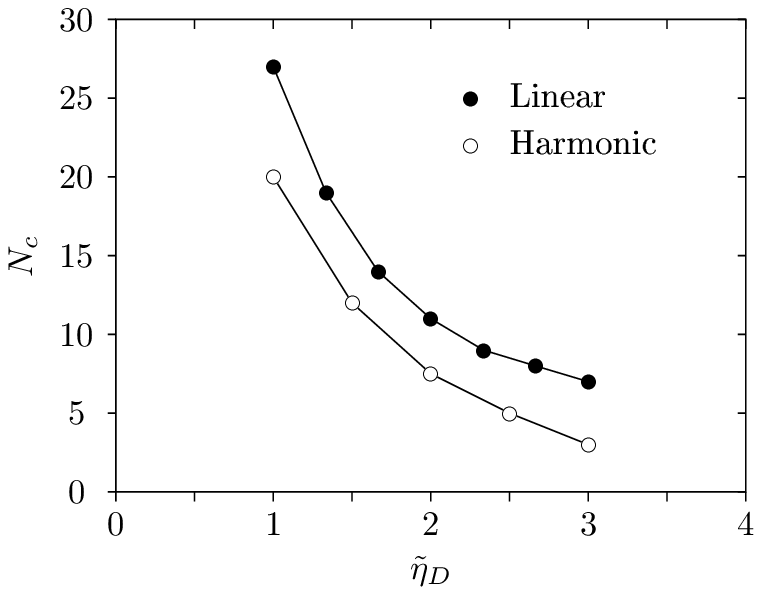}
\caption{Number $N_c$ of collisions necessary for rethermalization
between two evaporation zones (inferred by mapping the continuous
evaporation model onto the discrete one, see text) as a function
of the scaled evaporation parameter $\tilde{\eta}_D$. The solid
(resp. open) circles correspond to a linear (resp. harmonic)
confinement.} \label{Fig:ncol}
\end{figure}

We now turn to the kinetics of evaporation: for a given shape of
the potential (linear or harmonic) and a given value $\eta_D$ of
the evaporation parameter in the discrete model, we determine the
corresponding parameter $\eta_C(\eta_D)$ for the continuous model,
which leads to the same `evaporation trajectory' in the $(T,\Phi)$
plane. This mapping allows us to extract information on the
kinetics aspects of the rethermalization between evaporation zones
for the discrete models. More precisely, we infer the  number
$N_c$ of collisions required to rethermalize between successive
zones. For this purpose, we integrate over time the collision rate
from the continuous model, with an evaporation parameter
$\eta_C(\eta_D)$, between two points $(T_n,\Phi_n)$ and
$(T_{n+1},\Phi_{n+1})$ of the evaporation trajectory obtained with
the discrete model.

To allow for a quantitative comparison between linear and harmonic
confinements, we scale the evaporation parameter $\eta_D$ as in
Ref.~\cite{ketterleEvap} by defining $\tilde{\eta}_D=\eta_D/2$
(resp. $\tilde{\eta}_D=\eta_D/3$) for a harmonic (resp. linear)
confinement. Evaporation with a given normalized parameter
$\tilde{\eta}_D$ then yields approximately the same flux
reduction, independently of the shape of the confining potential.
On Fig.~\ref{Fig:ncol}, we have depicted $N_c$ as a function of
$\tilde{\eta}_D$, for both linear and harmonic confinements. As
expected, $N_c$ decreases with increasing $\tilde{\eta}_D$, since
the atomic distribution is less and less affected by the
evaporation, yielding a faster relaxation towards equilibrium. For
a linear confinement, 50\% to 100\% more collisions (as compared
to the harmonic one) are required for rethermalization between
evaporation zones. This dependence of the kinetics on the shape of
the confining potential is reminiscent of what is known for
thermalization of confined gas mixtures~\cite{anderlini}: the
rethermalization time is shorter in a homogeneous system than for
a trapped cloud. In a power-law trap of exponent $\delta$, the
rethermalization time decreases when $\delta$ increases, which
simply originates from the different scaling laws of the density
of states.

Therefore, two competing effects need to be considered when one
studies the whole evaporation process: in a linear potential, the
kinetics is slow but the gains in collision rate and in
phase-space density scale more favorably~\cite{epjd} than in a
harmonic confinement. In terms of the minimum number of collisions
required to achieve a given gain in phase-space density, those two
effects compensate each other. For instance, we find that at least
500 (resp. 630) collisions are necessary to gain a factor $5\times
10^7$ in phase-space density, in a purely harmonic (resp. linear)
potential, for an evaporation parameter $\eta_D\simeq4.5$ (resp.
$\eta_D\simeq5.5$). However, in terms of evaporation length, the
difference between harmonic and linear confinements is still
large, as \emph{runaway} evaporation can only occur in the latter
case for a two-dimensional evaporation. As an example, we consider
two beams with the same initial flux ($7\times10^9$ atoms per
second) and temperature (200~$\mu$K), with an initial phase-space
density $\rho_{\rm i}\simeq8\times10^{-7}$, propagating at 60~cm/s
either in a purely harmonic guide, either in a purely linear one.
For the former case, the initial collision rate is
$\gamma\simeq37$~s$^{-1}$, and slightly decreases to
$\gamma\simeq29$~s$^{-1}$ after evaporation to degeneracy. The
evaporation is performed at $\eta_C=6$, a value that minimizes the
evaporation length $L_{\rm ev}$, which reaches about 11~m. For a
linear confinement, although the initial collision rate is only
$\gamma\simeq19$~s$^{-1}$, its final value reaches $380$~s$^{-1}$
due to the runaway character of the evaporation. The total length
needed is only 6~m. Interestingly, the total number of collisions
that actually occurred within the beam is almost the same for both
confinements ($\sim 550$), as is the evaporation parameter
minimizing $L_{\rm ev}$.

\begin{figure}[t]
\includegraphics[width=8cm]{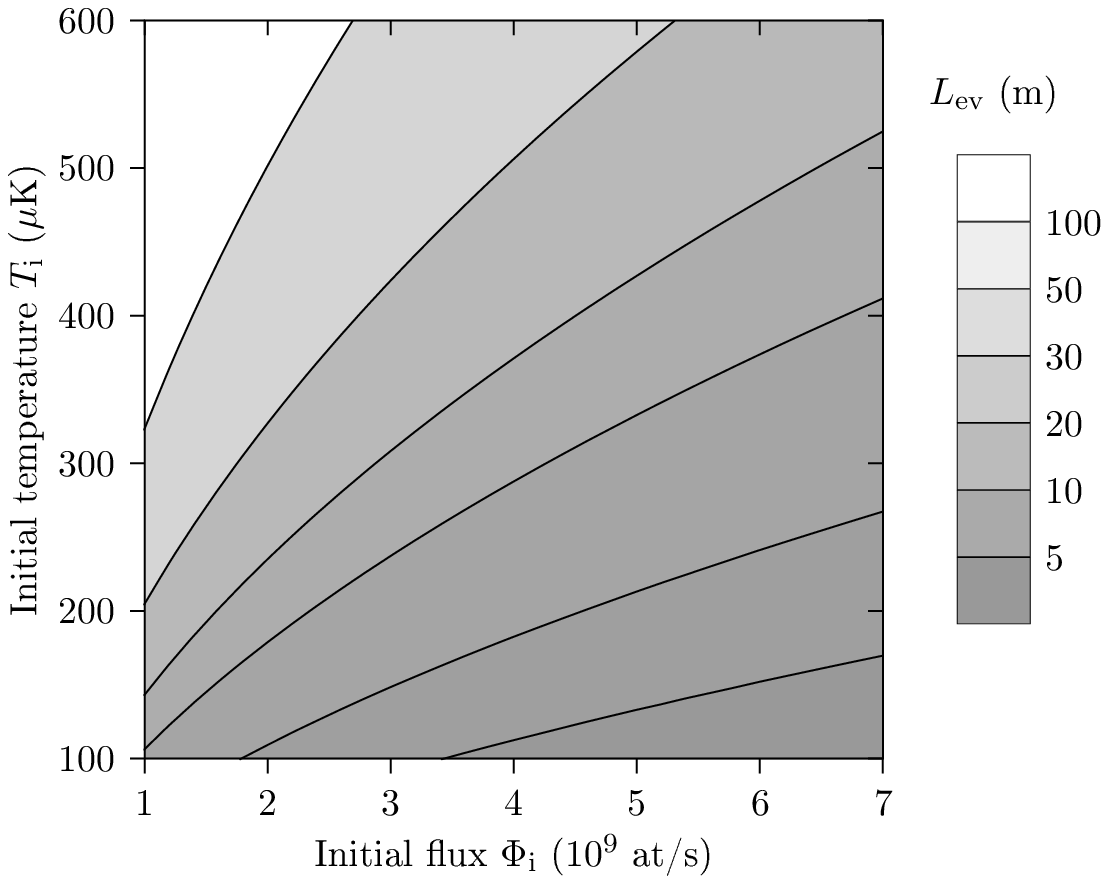}
\caption{Minimal evaporation length $L_{\rm ev}$ needed to reach a
phase-space density $\rho=1$ with evaporation at constant $\eta$
for an atomic beam of initial flux $\Phi_{\rm i}$ and temperature
$T_{\rm i}$, propagating at 60~cm/s in a guide with a gradient
$b=800$~G/cm and a longitudinal bias field $B_0=1$~G. The
iso-contours of $L_{\rm ev}$ show the large advantage of starting
with a temperature on the order of 100~$\mu$K for reaching
degeneracy in a reasonable length. The optimal evaporation
parameter $\eta$ is almost constant ($\simeq6$) over the whole
parameter space explored here.} \label{Fig:levap}
\end{figure}

In practice, due to Majorana spin-flips~\cite{lahaye05}, the guide
potential needs to be semi-linear. For a guide with parameters
$b=800$~G/cm and $B_0=1$~G, we have studied the minimal
evaporation length $L_{\rm ev}$ needed to reach degeneracy, as a
function of the initial flux $\Phi_{\rm i}$ and initial
temperature $T_{\rm i}$, assuming a beam velocity of $60$~cm/s.
The result is plotted on Fig.~\ref{Fig:levap}, and shows as
expected that $L_{\rm ev}$ decreases with lower $T_{\rm i}$ and
higher $\Phi_{\rm i}$. The optimal evaporation parameter is almost
constant with the value $\eta=6$ for our range of parameters
$(\Phi_{\rm i},T_{\rm i})$. The evaporation length determined this
way is very well fitted by a function of the form
\begin{equation}
L_{\rm ev}\simeq L_0\frac{T_{\rm i}^{3/2}}{\Phi_{\rm i}}\,.
\label{Eq:levap}
\end{equation}
This scaling can be easily understood: since here the runaway
effect exists only at the very beginning of the evaporation ramp,
before the effective shape of the potential becomes harmonic, the
evaporation length is simply inversely proportional to the initial
collision rate. For an initial flux $7\times10^9$~atoms/s, an
initial temperature $200$~$\mu$K, a guide gradient $b=800$~G/cm
and a bias field $B_0=1$~G, quantum degeneracy is reached for
$L_{\rm ev}\simeq7$~m, which shows the beneficial influence of the
increase of the collision rate at the beginning of the evaporation
ramp when the potential is essentially linear.

Actually, the evaporation length deduced here could be reduced by
lowering the beam's mean velocity (e.g, with the use of a tilted
guide~\cite{cnsns}) as it cools down, provided that the beam stays
supersonic enough.

\begin{acknowledgements}
We are indebted to Jean Dalibard for a careful reading of the
manuscript. We thank Johnny Vogels for stimulating discussions at
the early stage of this work. We acknowledge fruitful discussions
with the ENS laser cooling group, and financial support from the
D\'el\'egation G\'en\'erale pour l'Armement.
\end{acknowledgements}

\end{document}